\documentclass[12pt]{article}

\usepackage{amsmath,amssymb,bm,mathrsfs}
\usepackage{hyperref}
\usepackage{graphicx}

\newcommand{\beq}{\begin{equation}}
\newcommand{\eeq}{\end{equation}}
\newcommand{\beqn}{\begin{eqnarray}}
\newcommand{\eeqn}{\end{eqnarray}}

\newcommand{\beas}{\begin{eqnarray*}}
\newcommand{\eeas}{\end{eqnarray*}}

\newcommand{\bquo}{\begin{quote}}
\newcommand{\enqu}{\end{quote}}

\def\2{{1\over 2}}

\def\ba{\beq\new\begin{array}{c}}
\def\ea{\end{array}\eeq}

\begin{document}

\begin{titlepage}

\begin{flushright}
FTPI-MINN-16/07, UMN-TH-3517/16
\end{flushright}

\vspace{1cm}

\begin{center}
{  \Large \bf  Dynamically Emergent Flavor in a Confining \\[2mm]
Theory with  Unbroken Chiral Symmetry
}

\end{center}

\begin{center}
{\large
 Mikhail Shifman}
\end {center}

\vspace{1mm}

\begin{center}

{\it  William I. Fine Theoretical Physics Institute,
University of Minnesota,
Minneapolis, MN 55455, USA}

\end{center}

\vspace{0.6cm}

\begin{center}
{\large\bf Abstract}
\end{center}

I discuss ``truncated" QCD studied recently by Glozman {\em et al.} through numerical simulations. 
For two flavors it was observed that truncation restores the full chiral $U(2)\times U(2)$ symmetry of the Lagrangian. 
Moreover, additional enhancement of the above symmetry connecting representations with distinct Lorentz spins was observed. 

I argue that  the chiral symmetry restoration in a confining theory 
could entail emergent (extra) dynamical flavors which would show up  in the spectrum of color-singlet
particles provided their mass $\neq 0$. As an example,
I consider truncated QCD with a {\em single} massless Dirac quark. Assuming the validity of the above observations,
I demonstrate how a dynamical $SU(2)_{\rm fl}$ symmetry could emerge for {\em massive} spin-1 mesons without contradicting general principles.

\end{titlepage}

\section{Introduction}
\label{intro}

Spontaneous chiral symmetry breaking is a crucial feature of QCD. Since it occurs at strong coupling, so far there is no full understanding of the underlying dynamics. It was conjectured \cite{1p} that 
the chiral symmetry restoration could happen for highly excited (mesonic and baryonic) states, but this conjecture 
did not hold \cite{SVa}.

The standard QCD Lagrangian
\beq
{\cal L} = -\frac{1}{4g^2} G^{\mu\nu\,\, a} \,G_{\mu\nu}^a + \sum_{f}\bar\Psi iD_\mu\gamma^\mu\Psi
\label{one}
\eeq
can include one, two or more distinct Dirac fermions  -- flavors --  in the fundamental representation.  Denoting this number as $N_f$
we can say that with the vanishing quark masses  Lagrangian (\ref{one}) possesses a chiral symmetry
\beq
U(N_f) \times U(N_f)\,.
\label{two}
\eeq
 After spontaneous breaking of the chiral symmetry ($\chi$SB), and taking account of the axial anomaly, the flavor symmetry of conventional QCD is
\beq
SU(N_f)_{\rm diag}\times U(1)_V\,. 
\eeq
If $N_f=1$, i.e. with a single flavor, the conventional symmetry is just $U(1)_V$, the fermion charge.

The Banks-Casher formula \cite{BC} tells us that the quark condensate responsible for $\chi$SB does not develop provided the fermion Dirac operator modes  do not condense near zero (i.e. the density of the eigenvalues $\rho (0) =0$.) Thus, the near-zero modes play a special role in the chiral symmetry breaking. 

A conjecture was put forward 
in \cite{2p} that in lattice calculations with dynamical quarks in the vacuum but with {\em truncated} fermion Green functions, the chiral symmetry will be restored \cite{1,2}. By ``truncated" it is meant that  a (fixed) number of the lowest-lying eigenmodes of the Dirac operator are eliminated in the  quark propagator ``by hand." This framework will be referred to as ``truncated QCD." 

This conjecture was verified in recent calculations \cite{2p,1,2}.\footnote{See, however, Sec. \ref{prem} for a reservation.} It was found indeed that the chiral symmetry is restored in the spectrum, e.g. vector and axial vector mesons are degenerate. Moreover, an additional observation was that the effects due to the anomaly in the flavor-singlet
axial $U(1)$ current disappear. Thus, the first surprising ``experimental" finding of \cite{2p,1,2} is
that  the symmetry of the spectrum is at least $U(2)\times U(2)$  in truncated $SU(3)_{\rm c}$ QCD with  two flavors.\footnote{Here and below it will be assumed that all quark masses vanish.} 

It is clear that the above truncation introduces  nonlocal deformations in QCD. Their impact is unclear. For, instance,  it is not {\em a priori} certain that causality remains undamaged. Local  Lagrangian description in the ``truncation framework"  does not exist, and, moreover,  no non-local Lagrangian is known. We can be certain, however, that truncation does not break the Lorentz invariance. This is important. 

The second surprising finding of truncation is as follows.
The numerical results mentioned above exhibit an unexpected enhancement 
of the spectral symmetry, at least in the spin-1 sector -- the enhancement that goes beyond 
the expected  $U(2)\times U(2)$. A dynamical  $U(4)$ symmetry  was discussed  in the literature \cite{1pp,2pp}  (in truncated $SU(3)_{\rm c}$ QCD with  two flavors) in connection with this enhancement. 
The above $U(4)$ symmetry entangles geometric (dotted and undotted) spinorial quark indices with isospin, that would contradict the Coleman-Mandula theorem \cite{CM}, generally speaking. Below I discuss how a phenomenon looking as a ``flavor doubling" could emerge in the space of color-singlet mesons with mass $\neq 0$ in a hypothetical QCD with no $\chi$SB.

\section{Preliminaries}
\label{prem}

Let me try to concisely summarize main results of Glozman {\em et al.}.

Each massless 
Dirac fermion field in the fundamental representation of 
$SU(3)_{\rm c}$  is built from one dotted and one undotted Weyl spinor, $\chi_\alpha^i$ and $\bar{\eta}^{\dot\alpha\,i}$,  respectively, 
\beq
\Psi =\left(
\begin{array}{c}
\chi_\alpha^i\\[2mm]
\bar{\eta}^{\dot\alpha\,i}
\end{array}
\right),
\label{5}
\eeq
where $i$ is the color index (usually omitted in what follows). The standard definition of flavor implies that in the quark sector 
we have $N_f$ Dirac spinors (\ref{5}). 
The unbroken chiral $U(N_f)\times U(N_f)$ symmetry implies the following:

\vspace{1mm}

(i) If we denote the flavor indices $f$ and $g$ are introduced as $\chi^f$ and $\eta_g$,  
($f, g = 1,2,..., N_f$),  then all $2N_f$ fermion numbers, corresponding to the currents
\beq
\left(j_{\dot\alpha\alpha}\right)^f_f  = \bar\chi_{\dot\alpha\, f}\,\chi^f_\alpha\,,\quad \eta_{\alpha \, f}\, \bar\eta_{\dot\alpha}^f\,,
\quad \mbox{$f = 1, 2, ..., N_f$; no summation over $f$!}
\eeq
are conserved separately (see \cite{2p,1,2} in which $N_f=2$). 
In other words, there are $2N_f$ conserved quark charges. This follows from conservation of {\em all diagonal} vector and axial-vector currents, including the axial flavor singlet.

\vspace{1mm}

\vspace{1mm}

(ii) Conventional spin representations can be generalized to chiral spin: the hadronic states can be classified with regards to dotted and undotted indices separately, for instance, the state
$\bar\chi_{\dot\alpha}^f\,\chi^g_\alpha$ has  spin $S= ({\frac 12},  {\frac 12})$ while $\chi_\alpha^f\eta^\alpha_g$ has $S= (0,0)$.

\vspace{1mm}

(iii) Confinement of quarks in color-singlet states is not damaged.

\vspace{1mm}

(iv)  An additional degeneracy of the spectrum of the color-singlet mesons  connects 
chiral multiplets with interchanged dotted and undotted indices, for instance,
\beq
\chi_\alpha^i\, \eta_{\beta\,i} \leftrightarrow \bar{\eta}^{\dot\alpha\,i}\, \eta_{\beta\,i}\,.
\label{9p}
\eeq

\vspace{1mm}

(v) The least straightforward result of Glozman {\em et al.} is as follows. Despite the fact that the parity degeneracy
in the spin-1 sector is restored, which would normally imply that the $\chi$SB does not occur upon truncation (the biquark condensate does not develop), the fall off of the correlation functions of the scalar operators $\chi_\alpha\eta^\alpha$ and $\bar\chi_{\dot\alpha}
\bar\eta^{\dot\alpha}$ is not exponential, but rather power-like, implying that the spin-zero 
mesons are massless, as if they were Goldstones. In other channels (i.e. $J=1,2,...$) the corresponding mesons are massive.

\vspace{2mm}

 Needless to say, all of the above ``experimental" results should be  checked by independent group(s) before their status can be elevated to ``firmly established." A special emphasis should be put
 on clarification of the controversial point (v). 
 
 For the time being let us imagine, however,  that a consistent truncated version of QCD can be worked out and address the question
 whether an enhanced flavor symmetry can appear in the particle spectra upon the chiral symmetry restoration. Since this question can be raised even for one Dirac spinor\,\footnote{In this case the conjecture of \cite{2p,1,2} would imply $U(2)$ flavor symmetry instead of conventional 
 $U(1)\times U(1)$.} (i.e.  $N_f =1$) I will discuss namely this situation because of simplifying indices.

 \section{What happens if chiral symmetry is unbroken}
 \label{exp}
 
First, let us not that 
 two distinct ``diagonal" 2-point functions
\beq
 \left\langle \bar\chi_{\dot\alpha}\,\chi_\alpha (x) \,\, \left(\bar\chi_{\dot\alpha}\,\chi_\alpha (0)\right)^\dagger \right\rangle
 \quad \mbox{and}\quad 
  \left\langle \rule{0mm}{6mm}\chi_\alpha\eta_{\beta}  (x) \,\,  \left(\chi_\alpha\eta_{\beta}  (x) (0)\right)^\dagger \right\rangle
  \label{Dpp}
 \eeq
cannot be saturated by the same mesons. This is because a ``cross" correlator
 \beq
 \left\langle  \rule{0mm}{6mm}\bar\chi_{\dot\alpha}\,\chi_\alpha (x) \,\,  \left(\chi_\alpha\eta_{\beta}  (x) (0)\right)^\dagger \right\rangle
 \label{Dppp}
 \eeq
 vanishes identically which, in turn, follows from separate conservation of 
 both, the vector and axial current, i.e. 
 the $\chi$ and $\eta$ quark numbers.
 Thus, the correlation functions in (\ref{Dpp}) are saturated by different massive mesons. In the rest frame both have conventional spin $1$. However, the chiral spin structure 
 (which can be used for classification in the case of unbroken $\chi$SB is different. I also use the fact that 
 truncated QCD \cite{2p,1,2} does not violate Lorentz (in fact, Poincar\'e) invariance. 
The additional degeneracy of the spectrum of the color-singlet mesons detected by Glozman {\em et al.}  connects 
the following multiplets with interchanged dotted and undotted indices, for instance,
\beq
\chi_{\{ \alpha} \eta_{\beta\}} \leftrightarrow \bar{\eta}^{\dot\alpha}\, \eta_{\beta}\,
\label{9p}
\eeq
where the braces indicate antisymmetrization (three spin states) and 
\beq
\partial_{\dot\alpha}^\beta
\left(\bar{\eta}^{\dot\alpha}\, \eta_{\beta}\right)=0\,,
\label{10p}
\eeq
so that the operator on the right-hand side of (\ref{9p}) also produces three spin states (conventional spin 1).  The scalar-pseudoscalar state $\chi_{[ \alpha} \eta_{\beta]}$ has no partner because of (\ref{10p}), and the same for the $\chi$ current.
Since their ``experimental" interpretation is not yet clear, see above, I will not discuss them, focussing on
two different spin-1 states in (\ref{9p}).\footnote{In fact, each of the two operators produces two parity degenerate mesons.}

 In conventional QCD the symmetry of the theory under consideration is just vectorial $U(1)$. It is implemented trivially since all mesons have the corresponding charge zero.
 
 Restoration of the chiral symmetry in truncated QCD would lead to $U(1) \times U(1)$. Enhancement of symmetry needed for (\ref{9p}) was not expected {\em a priori}. We need something of the type of an $SU(2)$ converting $\chi$ into $\bar\eta$. If we do so
literally  we would brake the Lorentz symmetry by rotating an undotted spinor index into the dotted. We must act in a more subtle way. 

\section{Emergent dynamical flavor for spin-1 \\
mesons}
\label{df}

In our simplified example there are four color-singlet massive mesons
(if I choose interpolating fields without derivatives),
 \beq
 \chi^i_\alpha\eta_{\beta, i}\,,\quad   \bar\chi_{\dot\alpha\,i} \bar\eta_{\dot\beta}^i\,,\quad  \bar{\eta}^{\dot\alpha\,i}\, \eta_{\beta\,i}\,,\quad \chi_\alpha^i\, 
\bar{\chi}_i^{\dot\beta}\,.
\label{11}
\eeq
 There are four interpolating fields, each one produces generally speaking four chiral spin states. More exactly,
 we must focus only on the states with conventional spin 1, as was discussed above. This means that we must symmetrize with respect to $\alpha$ and $\beta$ in the first pair and take into account the fact that, due to transversality, (pseudo)scalar states are not produced by the second pair. Then each operator will produce three spin states. 
 
 A crucial point that for $M\neq 0$ where $M$ is the meson mass,
 each state (the Lorentz spins  $S= ({\frac 12},  {\frac 12})$ and  $S= (1,0)+(0,1)$),
 being distinctly different, can be described by one and the same formalism.\footnote{This was emphasized by Arkady Vainshtein.} Dotted indices can be converted into undotted and {\em vice versa} by applying the energy-momentum operator 
$ P_{\alpha\dot\alpha}$ or $ P^{\dot\alpha\alpha} $ which are invertible because $P^2=M^2\neq 0$.
Then, instead of the four operators (\ref{11}) with distinct Lorentz structure we can introduce
\beqn
 && \chi_\alpha\eta_{\beta}\,,\quad   \bar\chi_{\dot\alpha } \bar\eta_{\dot\beta}\,, 
 \nonumber\\[3mm]
&& \left( P_{\alpha\dot\alpha}\, M^{-1}\right)
\,  \bar{\eta}^{\dot\alpha}\, \eta_{\beta }\,,\quad 
\left(P^{\dot\alpha\alpha}\,M^{-1}\right) \chi_\alpha\, 
\bar{\chi}^{\dot\beta}\,.
\label{12}
\eeqn
I omitted the color indices as well as the symmetrization braces. Now, the operators in (\ref{12}) carry either both dotted or both undotted indices. It is important that $P_{\alpha\dot\alpha}$ is invertible.
$P_{\alpha\dot\alpha} P^{\dot\alpha\beta}=\delta^\beta_\alpha \, P^2$ and $P^2 >0$. Because of this fact, instead of (\ref{12}) one could also represent
all operators in the $S= ({\frac 12},  {\frac 12})$ form.

All four states in (\ref{12}) can now be related by standard $U(2)_{\rm fl} $ transformations.
Two diagonal transformations are equivalent to charge conservation for $\chi$'s and $\eta$' separately,
while off-diagonal ones are (see Appendix)
\beqn 
&&\delta \left(\chi^i_\alpha \eta_{\beta, i}\right) = \frac{i}{\sqrt{P^2}} \left[ P_{\alpha\dot\alpha} \left(\bar\eta^{\dot\alpha\,i}\eta_{\beta, i}\right)
+ P_{\beta\dot\beta} \left(\chi^i_\alpha \bar{\chi}_i^{\dot\beta}\right) \right]\bar\varepsilon \,,
\nonumber\\[2mm]
&&\delta \left(\bar\eta^{\dot\alpha\,i}\bar\chi^{\dot\beta}_{i}  \right) = \frac{i}{\sqrt{P^2}} \left[ P^{\dot\alpha\beta } \left(\chi^i_\beta 
\bar{\chi}^{\dot\beta}_{i}\right) +
 P^{\dot\beta\beta} \left(\bar\eta^{\dot\alpha\,i}\eta_{\beta, i}\right)   \right]\varepsilon \,,
\nonumber\\[2mm]
&&\delta\left(\bar{\eta}^{\dot\alpha\,i}\, \eta_{\beta\,i}\right)= \frac{i}{\sqrt{P^2}} \left[P^{\dot\alpha\alpha} \left(\chi_\alpha^i \eta_{\beta\,i}\right) \varepsilon +
 P_{\beta\dot\beta}\left( \bar\eta^{\dot\alpha\,i}\bar\chi^{\dot\beta}_i\right) \bar\varepsilon\,
\right],
\nonumber\\[2mm]
&&\delta \left(\chi_\alpha^i\, \bar{\chi}_i^{\dot\beta}\right)=  \frac{i}{\sqrt{P^2}} \left[P_{\alpha\dot\alpha} \left(\bar\eta^{\dot\alpha\,i} \bar\chi_i^{\dot\beta} \right)\bar\varepsilon\, +
 P^{\dot\beta\beta}\left(\chi_\alpha^i\eta_{\beta\,i} \right)\varepsilon\right],
 \label{13}
\eeqn
where $\varepsilon$ is a complex parameter of the off-diagonal transformations of $SU(2)_{\rm fl}$, and $\sqrt{P^2} = M$ when acting on a representation with a given mass $M$.  It is obvious that (\ref{13}) does not contradict the Coleman-Mandula theorem.
Convolution of two dotted or two undotted indices on the left-hand side of the first or the second line
will produce current divergences on the right-hand side, as I have already mentioned. 

It is obvious that the multiplet (\ref{12}) is closed, irreducible and all mesons in this multiplet must have degenerate masses. Altogether we have 12 distinct degenerate spin-1 states (including triple spin degeneracy), instead of two non-degenerate (real) sextets (including triple spin degeneracy) that would be present in the spectrum if the symmetry of the problem would be just $U(1)\times U(1)$.
In conventional QCD we would have four distinct real  spin-1  triplets. 

\section{\boldmath{$J=0$} mesons}

In the theory with one Dirac spinor under consideration, spin-zero mesons
are produced by the operators
$\chi_\alpha\eta^\alpha $ and $ \bar{\chi}_{\dot\alpha}\bar\eta^{\dot\alpha }$. In terms of real fields we have two degenerate mesons:
one scalar and one pseudoscalar. The parity degeneracy is due to the chiral symmetry restoration. Unlike spin-1 states,
no extra dynamical symmetry can emerge in this channel.

 \section{Two Dirac spinors}
 \label{twof}
 
 In principle, it is not difficult to generalize to two (or more) flavors. 
All enhanced degeneracies found by Glozman {\em et al.} can be explained by $U(2)_{\rm fl}$
 discussed in Sect. \ref{df} times   $SU(2)\times SU(2)$ chiral symmetry of theory with two Dirac spinors. 
 Whether or not full $U(4)$ emerges -- this will become clear only after the dynamical reason is fully understood.

 \section{Possible dynamical reason}
 \label{pdr}
 
 The symmetry as in (\ref{13}) could emerge in truncated QCD provided truncation suppress quark spin interactions
 in the background gluon field (to be integrated over in color-singlet two-point functions). Let me explain what I mean. In this (and only in this) section I will use the Dirac rather than the Weyl notation -- it turns out more economic in the case at hand.

Let us have a closer look at  the propagator of the
Dirac fermion in the operator form (see e.g. \cite{m4}), 
\beq
G(x,y)= \left\langle y \left|\,\, \slash \!\!\! \!{\cal P} \left({\cal P}^2 +\frac{i}{2} G_{\mu\nu} \,\sigma^{\mu\nu} 
\right)^{-1}\right| x\right\rangle
\label{6}
\eeq
where $G_{\mu\nu}$ is the background gluon field,
and ${\cal P}_\mu$ is the momentum operator in this background field. While ${\cal P}^2$ does not carry 
spinor indices, both $ \slash \!\!\! \!{\cal P}$ and $\sigma^{\mu\nu}$ have them. In particular,
\beq
G_{\mu\nu} \,\sigma^{\mu\nu} =  \left( 
\begin{array}{cc} 
\vec\sigma\vec{E} + 
i\vec\sigma\vec{B} & 0 \\[2mm]
0 &  -\vec\sigma\vec{E} + i\vec\sigma\vec{B}
\end{array}
\right),
\eeq
where  $\vec{E}$ and $\vec{B}$ are chromoelectric and chromomagnetic fields, respectively.\footnote{I use   gamma matrices in the spinorial representation.} 
The upper left corner is associated with propagation of $\chi$ while the lower right corner with $\bar\eta$. They do not mix in the massless theory. Usually a mass term is needed for infrared regularization of the Green function.
However, the truncation procedure discards the zero and low-lying modes and, therefore, automatically provides an infrared regularization.

Assume that truncation somehow suppressed the spin terms in (\ref{6}), so that one can replace (\ref{6}) by
\beq
G(x,y)\longrightarrow \left\langle y \left|\,\, \slash \!\!\! \!{\cal P} \, {\cal P}^{-2}  
\right| x \right\rangle\,.
\label{11ppp}
\eeq

Accepting (\ref{11ppp}) as a working hypothesis will result in a symmetry enhancement. 
In particular, the two-point functions in the channels $\left(\frac 12, \frac 12\right)$ and $(1,0),\,\, (0,1)$ 
become proportional. Namely
\beq
i\int d^4 x\, e^{iqx} \left\langle \bar\Psi (x)\Gamma \Psi(x)\,, \,  \bar\Psi (0) \Gamma \Psi(0)\right\rangle
\label{12pp}
\eeq
with $\Gamma =\gamma^\mu$ and $\Gamma =\sigma^{\mu\nu}$ can be shown to be equal up to trivial kinematical structures. 

Thus the assumption (\ref{11ppp}) would explain the degeneracy enhancement detected by Glozman {\em et al.}
The problem is that the very same assumption predicts even further degeneracy, between spin 1 and spin 0 (e.g. 
the correlation functions, say, for $\Gamma = \gamma^\mu$ and $\Gamma = \gamma^5$ are the same).
As was mentioned above the issue with scalar and pseudoscalar channels, as they are observed now
 in the numerical calculations of Golzman {\em et al.}, remains open, since 
the corresponding correlator cannot be fitted by exponentials.  So far, the exponential fit goes through
only  in the case of the spin-1, 2, etc. channels \cite{2p,1,2}. This situation is controversial.

\section{A curiosity: trading color for Lorentz indices}
\label{cur}
 
 This section is not directly related to the previous sections. It presents an observation in passing that
 was seemingly overlooked in the past.
 
 \subsection{\boldmath{$SU(4)_c$} and two-index antisymmetric quarks} 
\label{equ3}

This example was not discussed in
\cite{kogan} although the two-index antisymmetric  representation of $SU(4)_c$ is
quasi-real and should have been included in the
analysis. Since the spinors $\chi^{[ij]}_\alpha ,\,
\eta_{\alpha \, [km]}\varepsilon^{kmij}$ transform in the same manner under color and Lorentz transformation,
they can be rotated into each other. Thus the flavor symmetry in this case is $SU(2N_f)\times U(1)$. 
Since the Levi-Civita tensor $\varepsilon^{kmij}$ is symmetric under the interchange $[ij]\leftrightarrow [km]$
the pattern of the chiral symmetry breaking in (untruncated) QCD will be the same as for the adjoint quarks, namely,
\beq
SU(2N_f)\to SO(2N_f)\,,
\eeq
with  $2N_f^2+N_f-1$ Goldstone bosons (pions). For a single Dirac flavor we have two Goldstones, for two
Dirac  flavors nine pions in the symmetric two-index representation of $O(4)$. If the idea of \cite{2p,1,2}
-- the quark condensate suppression -- is correct then QCD truncation would eliminates all Goldstones and restore the full $SU(2N_f)$ flavor symmetry.

For $SU(3)_c$ two-index antisymmetric quark is identical to fundamental quarks.
The advantage of two-index antisymmetric quarks becomes obvious \cite{asv} at large $N$.

\subsection{\boldmath{$SU(2)_c$} and fundamental quarks} 
\label{equ2}

This example was analyzed long ago \cite{kogan}. In $SU(2)_c$ all representations are either
real or quasireal. The fundamental representation is quasireal.
The extended chiral symmetry at the Lagrangian level is $SU(2N_f)\times U(1)$ where
$N_f$ is the number of the $Dirac$ flavors. The pattern of the chiral symmetry breaking
in (untruncated) QCD is
\beq
SU(2N_f)\to Sp(2N_f)\,,
\label{7}
\eeq
with $2N_f^2-N_f-1$ Goldstone bosons (pions).
In the simplest example  $N_f=1$, the chiral symmetry is unbroken since $Sp(2)$
is isomorphic to $O(3)$ and to $SU(2)$. No Goldstones emerge. In the case $N_f=2$ considered in 
\cite{2p,1,2} we have $SU(4)\to Sp(4)\sim O(5)$, with five Goldstones. Suppressing the gluon condensate as in 
 \cite{2p,1,2} one can conclude that
in truncated QCD the full $U(4)_{\rm fl}$ is restored.

\subsection{Reducing to {\em D}=2 and 3}

Starting from the four-dimensional Dirac spinor (\ref{5}) in $SU(3)_{\rm c} $ QCD we can dimensionally reduce the theory down to $D=3$ or even $D=2$. More exactly, we discard one (or two) spacial dimensions leaving everything else intact.

If $D=3$ the number of the Lorentz rotations reduces to three: two boosts and one spacial rotation.
This implies that the Lorentz group is just a single $SU(2)$ and the distinction between dotted and undotted spinors disappears.

The Dirac spinor in  Eq. (\ref{5}) is composed of two two-component spinors; with two flavors we get four two-component spinors 
Then the flavor symmetry is obviously $U(4)$. 

The Lorentz group in two dimensions includes just a single boost. Needless to say, 
 the Dirac spinor  (\ref{5})   can be decomposed into two Dirac 2D spinors, again implying that the flavor symmetry is  $U(4)$. 

Thus, if four-dimensional truncated QCD dynamically ``selects" three-dimensional geometry at least with regards to spin degrees of freedom, then one would expect
the spectrum of the theory to be (approximately) $U(4)$ symmetric. By the same token, in  the general case of $N_f$ flavors we would get $U(2N_f)$. This is in no contradiction with the Coleman-Mandula theorem.

The above observations are 
summarized  in Figure \ref{one}.
\begin{figure}[htb]
 \center
  \includegraphics[scale=0.5]{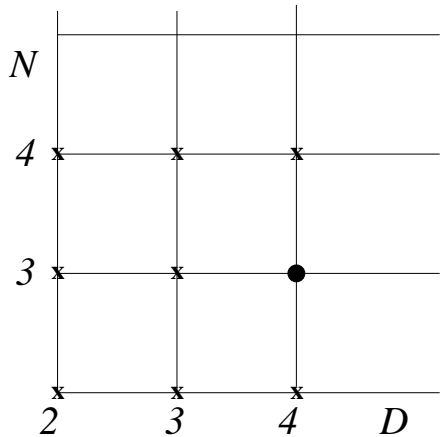}
  \caption{\small Chiral symmetry vs. number of dimensions and colors. Closed circle shows actual QCD. 
  Crosses denote theories with naturally enhanced chiral symmetry. }
\label{one}
  \end{figure}

\section{Caveats}

Truncated QCD with the vanishing  quark condensate (and presumed confinement) contradicts the Casher argument \cite{Casher}
that confinement necessarily leads to the chiral symmetry breaking. Although Casher's argument is imprecise,
it still tells us that the procedure of discarding near-zero modes from the quark propagators needs more
theoretical understanding.

\section*{Appendix}
\label{mne}
 
Here  I will describe  a mnemonic procedure (two generators) acting on quarks although it is ill defined for
 quarks because of their masslessnes. This Appendix is  only   for  mnemonics. 
 The actual physical context making it well-defined, is explained in Sect. \ref{df}.
It is important to understand that   the quark mass term or a non-vanishing quark 
 condensate would ruin the transformations to be presented below. However, we consider
 a massless quark in combination with vanishing biquark condensate in truncated QCD \cite{1p}.
 
 The energy-momentum operator $P_{\alpha\dot\beta}$ generating space-time shifts carries one dotted and one undotted index. Therefore the transition $\alpha\leftrightarrow\dot\alpha$
 can be achieved by combining $P_{\alpha\dot\beta}$ with the generators of $SU(2)\times SU(2)$ Lorentz rotations (boosts).  The meaning  of the operator $(P^2)^{-1/2}$  is explained in Sect. \ref{df}. The quark transformation ``laws"
 are
 \beq
 \delta\chi_\alpha = \frac{i}{\sqrt{P^2}}\,P_{\alpha\dot\alpha}\, \bar\eta^{\dot\alpha} \bar\varepsilon\,,\qquad
 \delta\bar\eta^{\dot\alpha}= \frac{i}{\sqrt{P^2}}\,P^{\dot\alpha \alpha}\chi_\alpha\,\varepsilon
 \eeq
 (and, of course,  Hermitean conjugate of the above). Here $\varepsilon$ is a complex transformation parameter. The $4\times 4$ generator matrices (analogs of the Pauli matrices) can be written as
\beqn
&&\Sigma_1 =\frac{1}{\sqrt{P^2}}\left(\begin{array}{cc}
0 & P_{\alpha\dot\alpha}\\[2mm]
P^{\dot\alpha\alpha}&0
\end{array}
\right),
\nonumber\\[3mm]
&&\Sigma_2 =\frac{1}{\sqrt{P^2}}\left(\begin{array}{cc}
0 & - i P_{\alpha\dot\alpha}\\[2mm]
i P^{\dot\alpha\alpha}&0
\end{array}
\right).
\label{15pp}
\eeqn
Moreover, the matrices
\beq
\Sigma_3 =\left(\begin{array}{cc}
I & 0\\[2mm]
0&- I
\end{array}
\right),\qquad {\boldmath\mbox{${ I}$}}=\left(\begin{array}{cc}
I & 0\\[2mm]
0& I
\end{array}
\right)
\label{15p}
\eeq
where $I$ is the $2\times 2$ unit matrix, being diagonal, are responsible for independent phase rotations of
We keep the above conservation laws in mind. The commutation relations for the generators (\ref{15pp})
are exactly the same as for the Pauli matrices.

\section*{Acknowledgments}

Illuminating discussions with L. Glozman and A. Vainshtein
are gratefully acknowledged. I would like to thank E. Shuryak for inviting me to the SCGP Conference
{\sl Gauge Field Topology: From Lattice Simulations and Solvable Models to Experiment},  (August  2015),
where this work was conceived. I am greatful to Adi Armoni for useful communications.

This work is supported in part by DOE grant DE-SC0011842.

\end{document}